\documentclass[conference]{IEEEtran}
\usepackage[T1]{fontenc}
\usepackage{cite}
\usepackage{amsmath,amssymb}
\usepackage{graphicx,booktabs,tabularx,array}
\usepackage{url}
\usepackage[hidelinks]{hyperref}
\newtheorem{proposition}{Proposition}
\newcommand{\LocalMedian}{0.234}
\newcommand{\LocalPninetyfive}{0.337}
\newcommand{\LocalPninetynine}{0.487}

\newcommand{\IWAP}{\textsc{IWAP}}

\newcommand{\Decode}{\mathsf{Decode}}

\newcolumntype{Y}{>{\raggedright\arraybackslash}X}
\hypersetup{
	pdftitle={Intent-Bound Authorization for Cryptocurrency Withdrawals Under Backend Compromise},
	pdfauthor={Toqeer Ali Syed, Asadullah Abdullah Khan, Ali Akrama},
	pdfsubject={IEEE-style research preprint; reference-model evaluation}
}

\title{Intent-Bound Authorization for Cryptocurrency Withdrawals Under Backend Compromise}

\author{
	\IEEEauthorblockN{
		Toqeer Ali Syed,
		Asadullah Abdullah Khan\textsuperscript{*},
		and Ali Akrama
	}
	\IEEEauthorblockA{
		Faculty of Computer and Information Systems\\
		Islamic University of Madinah, Saudi Arabia\\
		Emails: toqeer@iu.edu.sa, 471014356@stu.iu.edu.sa, 443059463@stu.iu.edu.sa\\
		\textsuperscript{*}Corresponding author: Asadullah Abdullah Khan
	}
}

\begin{document}
	\maketitle
	\begin{abstract}
		Binance's preliminary account of its October 2026 attack raises a custody question that key protection alone cannot answer: how can a provider prevent unauthorized withdrawals when an attacker controls a service permitted to request signatures? We propose the Intent-Bound Withdrawal Authorization Protocol (IWAP). Before signing, independent enforcement checks customer intent, exact transaction effects, fresh workload evidence, and a committed single-use spending reservation. Administrative actions use a separate authorization path. We specify the protocol's trust assumptions and give conditional arguments for intent preservation, replay resistance, and a fixed-epoch exposure bound. A Python reference monitor authenticates synthetic records with Ed25519 and serializes reservations with SQLite. It rejects 1,200 negative requests, accepts 3,000 valid synthetic requests, and preserves a 100-unit budget under 64 concurrent attempts in each of 20 trials. When the intent authority is deliberately compromised, malicious requests pass, but the independent budget still limits admitted exposure. The study evaluates a normalized policy model, not Binance's infrastructure or a deployed custody platform. Its contribution is an explicit authorization boundary and reproducible tests of that boundary.
	\end{abstract}
	
	\section{Introduction}\label{sec:intro}
	A cryptocurrency exchange can lose funds without an attacker extracting its private keys. If a compromised service can persuade protected signing infrastructure to authorize the wrong transaction, the resulting signature may be genuine and the transfer valid under blockchain rules. The missing property is the connection between the customer's instruction and the transaction that receives the signature. That connection must survive compromise of the component constructing the transaction.
	
	As of October 31, 2026, Binance's October 30 announcement reported unauthorized hot-wallet withdrawals beginning on October 29 and an estimated loss of approximately USD 6.3 billion. It stated that the investigation had found no evidence of leaked wallet private keys, while additional breach points were still being investigated \cite{binance2}. An earlier update suggested a possible route through a compromised DevOps laptop and a service container \cite{binance1}. These are preliminary statements, not a complete forensic record. We use the incident to motivate a backend-compromise threat model; we do not claim that this model reconstructs the attack or that the proposed protocol would necessarily have prevented it.
	
	Several familiar defenses leave this authorization question unresolved. A hardware-protected key can still sign an attacker-chosen message through an allowed API. A valid workload identity can belong to an exploited service. Attestation can establish claims about software state without establishing that a customer requested a payment. Blockchain verification can confirm settlement of an unauthorized transaction. An intrusion detector can miss the attacker or respond after signatures have been released. Effective custody requires an enforcement point that checks the particular withdrawal rather than inferring permission from these separate assurances.
	
	We propose \IWAP{}, a single authorization lifecycle for that enforcement point. An independent intent authority authenticates the customer's payment details. The backend constructs proposed transaction bytes but cannot create intent or change policy. Before a signature is released, independent enforcement verifies the exact effects, checks fresh workload evidence, and requires a durable reservation that consumes the intent and charges all applicable budgets. Wallet administration, delegation, and treasury replenishment follow separate governance rules. Behavioral alerts may restrict authority but cannot supply missing payment authorization.
	
	The design has two distinct objectives. With an honest intent source and honest enforcement in every usable signing quorum, backend compromise should not change the authorized payment. If the intent source is also compromised, intent correctness is lost; the remaining objective is a bounded quantity of admitted exposure while independent budget and governance boundaries remain intact. Stating both cases prevents the spending cap from being mistaken for a guarantee of zero loss under arbitrary compromise.
	
	The contribution is a protocol specification with conditional invariant arguments and reproducible tests of manipulated requests, concurrent admissions, and compromised-source intent.
	
	\section{Background and Related Work}\label{sec:related}
	\subsection{Identity and execution evidence}
	Zero Trust requires explicit access decisions rather than implicit permission based on location or ownership. NIST's cloud-native guidance extends that principle to application and service identities \cite{ztacloud}. This helps determine which workload may contact a signer. It does not establish that every payment proposed by that workload is authorized after compromise. IWAP therefore separates service access from transaction admission.
	
	The RATS architecture distinguishes an attester, verifier, and relying party, while the Entity Attestation Token standard defines attestation-oriented claims \cite{rats,eat}. These mechanisms support appraising an execution environment. Runtime work such as CFA+ adds evidence and protection concerning control-flow behavior \cite{cfaplus}. However, an allowed control-flow path can process malicious input. We use attestation to support the requesting and enforcing environments, while independently verifying the payment's meaning.
	
	\subsection{Permissions, signatures, and wallet control}
	Rich Authorization Requests provide structured authorization details, and DPoP binds OAuth tokens to proof of possession \cite{rar,dpop}. They inform narrow request permissions but do not, by themselves, implement customer balance reservation or one-use custody authorization. Badertscher et al.'s policy-compliant signatures connect signature validity to credential-policy satisfaction in anonymous payments \cite{pcs}. Their compliance setting differs from the transaction-construction and shared-budget problem examined here.
	
	Threshold signing distributes cryptographic authority. FROST specifies a two-round Schnorr protocol with defined ciphersuites \cite{frost}; Boneh et al. add proactive refresh to accountable threshold signatures \cite{refresh}. Such mechanisms protect participation in signing, but a quorum can still endorse the same malicious request if every participant trusts one compromised approval service. IWAP requires an honest independently verifying participant in every usable quorum. It does not implement or propose a replacement threshold-signature primitive.
	
	Wallet extensibility also changes the meaning of an approval. EIP-7702 authorizations designate code for an externally owned account, and ERC-7579 defines modular smart-account interfaces \cite{eip7702,erc7579}. A delegation or module change can alter future authority even when the immediate transaction transfers little value. A routine withdrawal profile must therefore distinguish payments from configuration changes, including changes nested inside batches or self-calls.
	
	\subsection{Detection and runtime enforcement}
	Provenance-based detection can help identify an intrusion before it reaches the signing path. MAGIC studies APT detection using graph representations, while ORTHRUS emphasizes actionable attribution \cite{magic,orthrus}. Their role differs from transaction authorization. An absent alert is not evidence that a recipient was approved, and an alert arriving after signature release may be too late to stop broadcast.
	
	Runtime guards provide a closer enforcement analogy. Chen et al. evaluate invariant guards against exploited smart contracts \cite{invariants}. That benchmark does not establish protection against a compromised exchange requesting ordinary valid transfers. IWAP defines custody-specific invariants instead: exact intent binding, persistent one-use state, separated administrative authority, and reserved exposure. The contribution lies in specifying their composition and failure semantics. We do not claim that transaction policies, attestation, or spending limits are new inventions.
	
	\subsection{Approval of displayed versus signed content}
	Approving what an interface displays rather than the bytes that are signed is a known failure pattern. In the February 2025 Bybit theft, attackers who had compromised a developer machine at the multisignature wallet provider injected malicious JavaScript into the wallet interface. Signers saw what appeared to be a routine transfer, but their approvals authorized a delegated call that replaced the wallet's implementation contract with attacker-controlled code, and more than USD 1.4 billion in assets was stolen \cite{bybitncc}. The multisignature cryptography itself was not broken, and the substituted operation is the kind that IWAP's operation profile reserves for a separate governance path. Regulated payments face an analogous requirement: the PSD2 regulatory technical standards require that the authentication code be specific to the amount and payee and be invalidated if either changes, a requirement known as dynamic linking \cite{psd2rts}. IWAP applies the same principle to custody, but binds intent to the decoded effects of the exact bytes at independent signers rather than to a display layer, and adds single-use and budget state that dynamic linking does not address.
	
	\section{Problem and Trust Assumptions}\label{sec:model}
	The primary adversary controls the withdrawal backend, including transaction construction, request scheduling, ordinary service credentials, and local state. It may replay certificates, substitute transaction fields, race requests across networks, withhold broadcasts, and interrupt communication. A legitimate workload identity may remain valid. The adversary cannot be assumed to announce its presence through an anomaly detector.
	
	Five assumptions define the proposed protection. First, an independent intent authority authenticates transaction-specific customer confirmation and accurately certifies a reserved balance. Second, every quorum capable of releasing a chain signature includes honest enforcement of the admission rule. Third, the state authority provides durable, serializable reservations and authentic commitment evidence. Fourth, a sound chain adapter interprets the exact signed bytes and all relevant effects. Fifth, separate governance protects policy versions, budget increases, signer membership, recovery, and replenishment. Cryptographic primitives and the chosen hardware roots are trusted only under their stated security models.
	
	These assumptions require administrative separation. Shared privileged credentials, approval APIs, or recovery secrets can collapse nominally separate services into one failure domain. A fully compromised signing quorum can bypass off-chain policy; on-chain guards help only if the adversary cannot bypass or remove them.
	
	We define an unauthorized withdrawal as one whose economic or control effects exceed the authenticated instruction and policy. This includes recipient, amount, asset, chain, fee, or operation substitution; repeated economic admission; and aggregate budget violations. Consensus attacks, arbitrary DeFi exploits, bridge failures, and customers knowingly approving an attacker's address are outside the prevention claim. They require different controls or additional transaction profiles.
	
	\section{The IWAP Protocol}\label{sec:protocol}
	\subsection{Independent intent and exact effects}
\begin{figure*}[t]
		\centering
		\includegraphics[width=\textwidth]{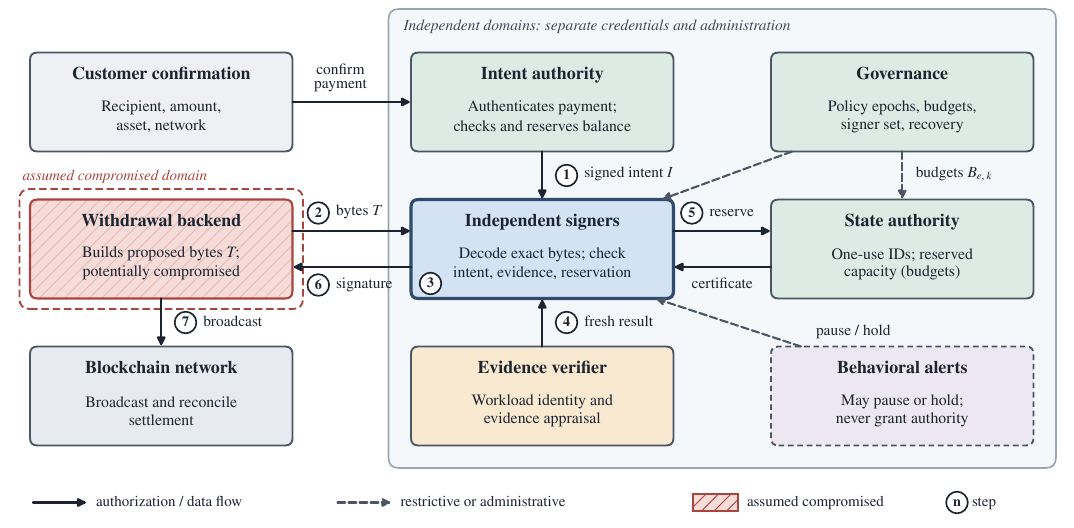}
		\caption{IWAP's logical boundaries. The backend constructs transactions but supplies no independent financial authority. Signers require intent, a committed reservation, and appraised workload evidence. The distributed architecture is specified; the experiment implements a local admission subset. Numbers follow the admission flow in Section~\ref{sec:flow}.}
		\label{fig:architecture}
	\end{figure*}

	Figure~\ref{fig:architecture} shows the independent roles. An intent is a structured record
	\begin{equation}
		I=(r,c,w,n,a,d,q,f,\nu,o,e,t_{\rm exp},b),
		\label{eq:intent}
	\end{equation}
	where $r$ identifies the request, $c$ the customer, $w$ the wallet scope, $n$ the network, $a$ the asset, $d$ the recipient, $q$ the amount in atomic units, $f$ the fee ceiling, $\nu$ the nonce or input commitment, $o$ the operation, $e$ the policy epoch, $t_{\rm exp}$ the expiry time, and $b$ the balance-reservation reference. A production encoding also fixes a schema version and network-specific auxiliary fields. The intent authority signs a canonical, domain-separated commitment after authenticating the customer's confirmation. A generic login or multifactor prompt that does not bind payment details is insufficient.
	
	The backend proposes bytes $T$. An adapter derives $\Decode_n(T,\chi_n)$ using verified chain context $\chi_n$. Independent enforcement compares all relevant effects with $I$ and the supported operation profile. A native transfer, token transfer, UTXO transaction, and instruction-based transaction need different parsers. UTXO change, fee computation, token behavior, auxiliary recipient data, and additional instructions cannot be ignored. Unknown or ambiguous effects cause a hold.
	
	The signer reconstructs the digest from the verified bytes. It does not trust an unexplained digest supplied by the backend. The signing session binds that digest to the intent identifier, reservation, epoch, and signer set. A later change requires another decision. Routine withdrawal credentials cannot approve delegation, module installation, ownership transfer, upgrades, or arbitrary token approvals. Those operations require a separate governance path even if their immediate transferred value is zero.
	
	\subsection{Admission flow}\label{sec:flow}
	A withdrawal passes through the following ordered steps; the backend supplies the proposed bytes in step 2 and only receives the signature and broadcasts in steps 6 and 7.
	\begin{enumerate}
		\item The customer confirms recipient, amount, asset, and network. The intent authority signs $I$, certifies the balance reservation $b$, and delivers $I$ directly to the independent enforcement participants.
		\item An enforcement participant verifies the intent signature, decodes $T$, and evaluates $\mathsf{ExactMatch}$, $\mathsf{OperationAllowed}$, $\mathsf{EpochCurrent}$, and $\neg\mathsf{Paused}$. Any failure or unknown effect ends the flow with a hold and no state change.
		\item The participant obtains a verifier result bound to the intent digest, the transaction digest, and a fresh challenge, establishing $\mathsf{EvidenceFresh}$.
		\item The participant, not the backend, asks the state authority for a reservation. The state authority atomically consumes the intent identifier, commits the digest, and charges every applicable bucket under (\ref{eq:budget}), then returns an authenticated certificate after durable commit.
		\item Each signer verifies the certificate against its own decoded digest, epoch, and current policy before releasing a share or signature, and the resulting signature is returned to the backend.
		\item The backend assembles and broadcasts the signed transaction.
	\end{enumerate}
	Inexpensive deterministic checks precede reservation so that invalid requests consume no capacity. If signing aborts after step 5, the charge remains until resolved as described in Section~\ref{sec:uncertainty}; an aborted signing session does not refund capacity.
	
	\subsection{Workload evidence with a limited role}
	A verifier result binds an approved workload profile to the intent digest, proposed transaction digest, epoch, challenge, and a short acceptance interval. Its signature authenticates the verifier's appraisal. It does not transform platform integrity into evidence of customer consent. Freshness limits replay of old results but does not eliminate runtime compromise between appraisal and signing.
	
	Propositions~1--3 below do not invoke this result directly. Its role is to support the second trust assumption, by showing that the enforcing environments run approved builds, and to constrain the requesting workload, so that a compromised or unapproved build is held before it can request signatures. It is a necessary condition in (\ref{eq:admit}), never a sufficient one.
	
	Figure~\ref{fig:architecture} places this result alongside independent financial evidence. Behavioral or APT alerts can pause an affected workload or request review. A low-risk score cannot override mismatched payment details or an exhausted budget. This restrictive direction keeps the detector out of the authority-granting path and preserves the deterministic checks when detection misses an attack.
	
	\subsection{Reserve exposure before signing}
	For each policy bucket $k$, let $B_{e,k}$ be the capacity allocated in epoch $e$ and $U_{e,k}$ the committed charge. A request requiring nonnegative charge $x_k$ is admitted only if
	\begin{equation}
		U_{e,k}+x_k\leq B_{e,k}\quad\text{for every applicable }k.
		\label{eq:budget}
	\end{equation}
	Checking the one-use identifier, consuming it, committing the exact digest, and updating every bucket occur atomically. The state authority returns an authenticated certificate after durable commit. Signers verify that certificate before releasing shares or signatures. Multiple customer, wallet, asset, and provider-wide limits may apply, but they form one admission decision.
	
	Cross-chain protection concerns shared authorization capacity, not atomic execution on different blockchains. Let $B_e$ denote the provider-wide limit in epoch $e$. A provider can serialize this common budget or allocate disjoint per-network capacities $B_{e,m}$ satisfying $\sum_m B_{e,m}\leq B_e$, where $m$ ranges over networks. Each region or chain must not receive the entire global limit independently. Reallocation must prevent the same capacity from being granted twice during delay or failure.
	
	The mandatory admission conditions can be written as
	\begin{equation}
		\begin{aligned}
			\mathsf{Admit}={}&\mathsf{IntentValid}\land\mathsf{ExactMatch}\\
			&\land\mathsf{OperationAllowed}\\
			&\land\mathsf{EvidenceFresh}\land\mathsf{EpochCurrent}\\
			&\land\mathsf{ReservationCommitted}\\
			&\land\neg\mathsf{Paused}.
		\end{aligned}
		\label{eq:admit}
	\end{equation}
	Unavailable mandatory evidence produces a hold. An authentic certificate can still be expired, associated with another digest, or contrary to the current policy. Successful signature verification alone never satisfies the complete predicate.
	
	\subsection{Uncertainty, revocation, and replenishment}\label{sec:uncertainty}
	A committed reservation remains charged after signature release, broadcast uncertainty, and settlement within the fixed epoch. A timeout does not prove that a signature or transaction is unusable. Retries must preserve the original economic commitment, while replacement transactions need an explicit equivalence and exposure rule. Refunding capacity merely because one RPC endpoint reports no transaction can admit duplicate exposure.
	
	Off-chain expiry governs new signing admission. It does not universally cancel a signature already released, especially for ordinary account or UTXO transfers. Revocation likewise blocks new releases through honest enforcement without erasing existing signatures. Reservations can be resolved only through a defined settlement or chain-specific invalidation condition. New policy epochs carry forward unresolved exposure or account for it in their newly granted capacity.
	
	Governance maintains a monotonic policy history and controls changes that increase authority. Restoring an old disk snapshot must not restore an old signing policy. Treasury refills also need independent authorization; otherwise repeated replenishment defeats an apparently small hot-wallet cap. Emergency recovery follows the same separation: a pause function must not quietly become an unrestricted fund-transfer endpoint.
	
	\section{Conditional Security Properties}\label{sec:security}
	The protocol supports three conditional properties. The arguments concern the specification and its assumptions; they are not mechanized proofs of a chain parser, threshold implementation, or distributed state service.
	
	\begin{proposition}[Intent preservation]
		If the intent authority is honest, transaction decoding is sound, and every usable signing quorum contains honest enforcement, backend compromise cannot produce a release outside the authenticated payment's permitted effects, except through failure of the assumed cryptographic primitives.
	\end{proposition}
	\begin{IEEEproof}
		An honest enforcing participant verifies the independently authenticated intent and decodes the exact bytes to be signed. A changed recipient, amount, network, or operation fails matching. A digest changed after checking fails session binding. A successful unauthorized release therefore requires forged intent, an unsound adapter, a signing bypass, or a fully compromised enforcement quorum, contrary to the assumptions.
	\end{IEEEproof}
	
	\begin{proposition}[Single economic admission]
		Durable, serializable consumption of an intent identifier admits at most one transaction commitment for that identifier.
	\end{proposition}
	\begin{IEEEproof}
		Concurrent reservations are serialized. At most one observes the identifier as unconsumed and atomically inserts its digest commitment. Later attempts are rejected or return the same idempotent commitment without new capacity. A crash cannot remove the durable association under the state assumption.
	\end{IEEEproof}
	
	\begin{proposition}[Fixed-epoch exposure]
		With nonnegative charges, retained reservations, honest budget enforcement, and no unaccounted capacity increase, $U_{e,k}\leq B_{e,k}$ holds for every reachable state.
	\end{proposition}
	\begin{IEEEproof}
		The initial state satisfies the cap. An admission increases the charge only after checking (\ref{eq:budget}) atomically. Rejection leaves state unchanged. Signing, uncertainty, and finalization retain the charge. Induction over these transitions preserves the invariant.
	\end{IEEEproof}
	
	The last property remains meaningful after intent-authority compromise, but then it bounds admitted exposure rather than establishing legitimate intent. It is also limited to its stated horizon and accounting units. A new budget every hour can accumulate losses over hours. A common fiat-valued budget additionally needs conservative asset valuations, fee accounting, and a policy for price-envelope failures. The experiment uses synthetic units and makes no market-valued loss claim.
	
	\section{Implementation and Evaluation}\label{sec:eval}
	\subsection{Reference monitor and reproducibility}
	The reference monitor is implemented in Python 3.12.14 using the cryptography package (version 50.0.2) for software Ed25519 signatures and SQLite 3.53.1 through Python’s sqlite3 module. The program issues no PRAGMA statements, leaving journal and synchronization settings at their defaults. Monitor connections use a 30-second busy timeout and explicitly controlled transactions. All database-backed experiments use an in-memory database except the concurrency test, which uses an on-disk file. Separate test issuers authenticate synthetic intent and workload-result records. With all controls enabled, the monitor requires the transaction to match the signed intent on twelve normalized fields, corresponding to the intent fields in Eq.~(\ref{eq:intent}) except the balance-reservation reference $b$, which this implementation does not model. It also checks a positive integer amount, a nonnegative integer fee ceiling, transfer-only operation, the current epoch, intent expiry, the evidence signature, request hash, workload identity, approved-build identifier, and a 30-second synthetic freshness interval. Finally, it uses BEGIN IMMEDIATE to serialize intent-identifier consumption and reserve the principal amount plus the fee ceiling against the spending budget.
	
	This subset does not implement customer authentication, balance-reservation proofs, hardware quotes, runtime instrumentation, real transaction parsers, multiparty signing, or blockchain settlement. Its chain names are labels on generated records. The full protocol requires stronger schema validation and domain-separated production encodings. The test artifact authenticates restricted ASCII JSON records and is not a hostile-input custody API.
	
	The negative suite contains twelve scenario families with 100 generated instances per family, using seeds 0--99 for small payment amounts. A fresh monitor isolates each case; replay cases first consume the intent. The families cover recipient, amount, and network substitution; delegation disguise; an untrusted intent issuer; duplicate admission; expiry; an old epoch; stale evidence; evidence for another request; an unapproved measurement; and excess capacity. The raw output records each decision and rejection reason.
	
	A separate workload submits 3,000 valid requests over 30 seeds. A fresh in-memory database and ample budget for each seed avoid intentional capacity failures. Local timing surrounds certificate verification, policy checks, and reservation; evidence issuance, networking, hardware operations, signing, and settlement are excluded. The reported percentiles describe one recorded host run (x86\_64 Linux, kernel 6.18.57, nine logical CPUs) rather than a provider service-level objective.
	
	\subsection{Concurrency, compromise, and control removal}
	The concurrency experiment sends 64 distinct requests, each charging 19 units of principal and a one-unit fee ceiling, against a shared 100-unit budget. Sixteen worker threads use separate connections to an on-disk SQLite database. Twenty repetitions each begin with fresh state. Alternating abstract network labels test one shared counter, not real cross-chain execution or distributed consensus.
	
	A stronger-compromise experiment gives the adversary the test intent issuer's key and valid workload evidence. It submits 100 nine-unit payments with one-unit fee ceilings under separate caps of 100, 250, and 500 units. This deliberately removes intent-source trust while preserving independent budget enforcement. We also treat each check as a mutation-testing target: seven targeted ablations separately remove binding, operation restriction, replay prevention, budget checks, epoch checks, expiry checks, or evidence appraisal. Each mutant receives a request constructed to exploit that omitted condition, which tests whether the suite exercises each check independently.
	
	Finally, four 1,000-request runs inject stale evidence into 0\%, 1\%, 5\%, or 10\% of otherwise valid requests. A small abstract explorer separately traverses two intents with costs one and two, a budget of three, and one revocation. Its states distinguish unreserved, reserved, signed, uncertain, and finalized requests. Neither this traversal nor the fault injection evaluates a learned detector.
	
	\subsection{Observed results}
	Table~\ref{tab:results} summarizes the executed reference study. All 1,200 negative requests were rejected, and all 3,000 valid synthetic requests were admitted. The local admission median was \LocalMedian{} ms, with 95th and 99th percentiles of \LocalPninetyfive{} and \LocalPninetynine{} ms. These observations establish conformance for the tested records. The selected negative cases are not a statistical sample of attacks, so zero admissions does not estimate a universal prevention rate.
	
	\begin{table}[t]
		\caption{Executed reference-model results and their scope}
		\label{tab:results}
		\centering\footnotesize
		\begin{tabularx}{\columnwidth}{p{.17\columnwidth}YY}
			\toprule Experiment & Observation & Scope limit\\\midrule
			Negative suite & 0 of 1,200 generated negative requests admitted. & Selected cases; not a prevention rate.\\
			Valid workload & 3,000 of 3,000 generated valid requests admitted. & Synthetic records; local timing only.\\
			Concurrent budget & 5 of 64 requests admitted; 100 units reserved in every one of 20 trials. & One host; no partitions or Byzantine replicas.\\
			Compromised issuer & Principal admitted: 90, 225, 450; reserved charges: 100, 250, 500. & Synthetic units, not blockchain losses.\\
			Targeted ablations & All seven omissions admitted their corresponding constructed bypass. & Constructed requests; no attack frequency.\\
			Stale evidence & 0, 10, 50, 100 legitimate requests held in the respective 1,000-request runs. & Availability cost, not a detector false-positive rate.\\
			Abstract exploration & 50 states; 115 generated transitions; no encoded invariant violation. & No refinement or liveness claim.\\\bottomrule
		\end{tabularx}
	\end{table}
	
	Each concurrency repetition admitted exactly five requests and reserved exactly 100 units. This is consistent with serializable pre-signing capacity allocation. It does not validate network partitions, Byzantine replicas, policy rollback, or regional capacity transfers. Those scenarios require a fault model and implementation beyond the local database used here. Multi-process contention, process crashes during a transaction, and comparison against a sequential reference model are not yet tested and are left to future work.
	
\begin{figure}[t]
		\centering\includegraphics[width=\columnwidth]{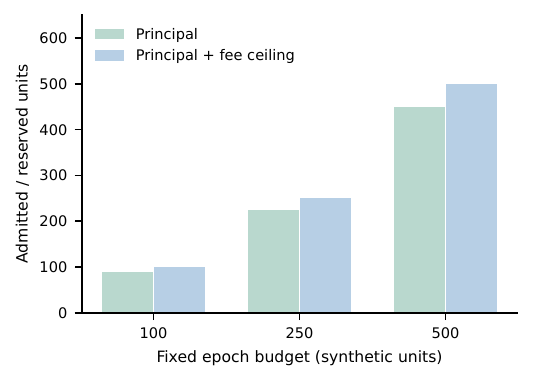}
		\caption{Compromised intent authority with an honest budget monitor. Each payment reserves nine units of principal and a one-unit fee ceiling.}
		\label{fig:bound}
	\end{figure}

	The compromised-issuer result in Fig.~\ref{fig:bound} exposes the boundary of prevention. Apparently valid malicious intent passes exact matching because its trusted source has failed. Nevertheless, the honest monitor limits total reservations to the configured cap. Accepted principal is below reserved exposure because each payment also reserves its fee ceiling.
	
	Every targeted control removal admitted its constructed bypass. Removing binding allowed a changed recipient; removing one-use state allowed duplicate admission; removing the operation rule admitted an authentically signed but forbidden delegation request. The remaining omissions similarly exposed their respective epoch, expiry, evidence, or capacity condition. These examples show that the checks enforce different requirements, not that any one removal has a measured real-world attack frequency.
	
	Stale evidence caused exactly the injected number of holds. This demonstrates the availability cost of the chosen fail-closed rule, not a detector false-positive rate. The abstract explorer found no violation of its retained-charge, cap, and revocation-pause invariants across 50 states and 115 transitions. It does not prove refinement between the model and implementation or establish liveness under arbitrary failures.
	
	\section{Deployment Implications and Limitations}\label{sec:discussion}
	A provider evaluating IWAP should first enumerate every route capable of producing a usable signature. Normal withdrawals, raw-digest APIs, backup recovery, vendor maintenance, emergency actions, and wallet upgrades all belong in that inventory. Adding a protected normal path while retaining an unrestricted fallback would leave the central assumption false. Administrative separation must also survive shared dependencies and recovery procedures.
	
	Deployment can proceed through transaction-intent capture, shadow semantic checks, and then enforced reservations at independent signers. Shadow checks help find legitimate edge cases but provide no prevention while passive. Before enabling an asset, the provider should test its adapter against recipient substitution, forbidden authority changes, fee manipulation, batching, retries, and uncertain broadcasts. Budget increases and treasury refills require separate review so that operational convenience does not silently expand an attacker's capacity.
	
	Two practical issues shape this deployment. First, exchanges commonly batch many withdrawals into one transaction, whereas the intent record in (\ref{eq:intent}) describes a single recipient and amount. A batch profile must match every decoded output to a distinct authenticated intent, charge the sum of principals and the transaction fee across all applicable buckets in one atomic reservation, and hold the whole transaction if any output, change address, or instruction has no matching intent. Second, a single provider-wide budget serializes all admissions and becomes a contention point. Allocating disjoint sub-budgets per wallet, region, or network reduces contention, but only under the reallocation rule given earlier; the local experiment does not measure this tradeoff.
	
	The reference study leaves substantial integration work. It lacks real chain adapters, hardware-backed evidence, an implemented threshold-signing quorum, replicated authorization state, and a protected customer-confirmation channel. Its negative suite exercises explicit requirements rather than adaptive adversaries against deployed code. Its timings exclude the stages likely to dominate a distributed signing path. A demanding practical-security claim would require those components and independent adversarial review.
	
	The Binance incident provides motivation, not demonstrated prevention. Its preliminary announcements do not establish every attacker privilege or affected signing route. Later forensic findings may change the relevant threat model. The synthetic study is not an incident replay and supplies no evidence about Binance's internal architecture.
	
	\section{Conclusion}\label{sec:conclusion}
	The October 2026 Binance attack motivates examining what authorizes a withdrawal after service compromise. IWAP places that decision at an independent boundary that checks payment intent, exact effects, workload evidence, and committed spending state before signing. Its conditional arguments separate intent preservation from exposure containment, and its reference experiments show both rejected manipulations and nonzero residual admission after intent-source compromise. The next step is to implement and test the same boundaries in real parsers, signing infrastructure, customer confirmation, and fault-tolerant state services. The current evidence supports a reproducible policy reference study, not a claim of deployed exchange protection.

	\section*{Reproducibility and Scope}
	
	The accompanying source package includes the evaluator, recorded results, raw negative outcomes, figure-generation scripts, and reference-verification notes. All experiments were conducted locally using synthetic records and generated test keys. No funds were transferred, and no production service was tested. The research artifacts are publicly available in the GitHub repository at \url{https://github.com/Asadullah779-pro/IWAP-Withdrawal-Authorization} and permanently archived on Zenodo as version v1.0.0, with DOI: \url{https://doi.org/10.5281/zenodo.23825420}.

	\bibliographystyle{IEEEtran}
	\bibliography{references}
\end{document}